\documentclass[aps,twocolumn,amsmath,amssymb,preprintnumbers,superscriptaddress,floatfix]{revtex4}

\usepackage{physics}

\usepackage[utf8]{inputenc}
\usepackage{newtxtext}
\usepackage[upint]{newtxmath}
\usepackage{microtype}
\usepackage{textcomp}
\usepackage{dsfont}
\usepackage{eucal}
\usepackage{siunitx}
\usepackage{soul}

\usepackage{enumerate}
\usepackage{amsfonts}
\usepackage{color}
\usepackage{soul}

\usepackage{todonotes}
\presetkeys%
    {todonotes}%
    {inline}{}

\usepackage{graphicx}

\usepackage[colorlinks,allcolors=blue]{hyperref}
\usepackage{cleveref}

\newcommand{\me}[1]{\mathrm{e}^{#1}}                            % Differential scalar element
\newcommand{\cc}[1]{{#1}^*}                                     % Complex conjugate
\newcommand{\transpose}[1]{\ensuremath{#1}^{\intercal}}

\definecolor{DarkBlue}{rgb}{0,0,0.80}
\definecolor{DarkRed}{rgb}{0.80,0,0}
\definecolor{Purple}{rgb}{0.55,0,0.55}
\definecolor{Purple}{rgb}{0,0,0.8}

\DeclareMathOperator{\diag}{diag}
\DeclareMathOperator{\antidiag}{antidiag}
\DeclareMathOperator{\rect}{rect}

\newcommand{\thouless}{\varepsilon_{\textsc t}}

\let\epsilon\varepsilon

\begin{document}
\title{Combined Zeeman and orbital effect on the Josephson effect in rippled graphene}

\author{Eirik Holm Fyhn}
\email{eirik.h.fyhn@ntnu.no}
\affiliation{Center for Quantum Spintronics, Department of Physics, Norwegian \\ University of Science and Technology, NO-7491 Trondheim, Norway}
\author{Morten Amundsen}
\affiliation{Center for Quantum Spintronics, Department of Physics, Norwegian \\ University of Science and Technology, NO-7491 Trondheim, Norway}
\author{Ayelet Zalic}
\affiliation{The Racah Institute of Physics, The Hebrew University of Jerusalem, Jerusalem 91904, Israel}
\author{Tom Dvir}
\affiliation{The Racah Institute of Physics, The Hebrew University of Jerusalem, Jerusalem 91904, Israel}
\author{Hadar Steinberg}
\affiliation{The Racah Institute of Physics, The Hebrew University of Jerusalem, Jerusalem 91904, Israel}
\author{Jacob Linder}
\affiliation{Center for Quantum Spintronics, Department of Physics, Norwegian \\ University of Science and Technology, NO-7491 Trondheim, Norway}

\date{\today}
\begin{abstract}
  \noindent The two-dimensional nature of graphene Josephson junctions offers the possibility of creating effective superconductor-ferromagnet-superconductor junctions with tunable Zeeman splitting caused by an in-plane magnetic field.
  Such junctions would be able to alternate between a conventional superconducting ground state and a ground state with an intrinsic phase difference, making them controllable \mbox{0-$\pi$} Josephson junctions.
  However, in addition to the Zeeman splitting, an in-plane magnetic field will in general also produce an orbital effect because of height variations in graphene, colloquially known as ripples.
  Both the Zeeman and orbital effect will thus affect the critical current, so to be able to identify \mbox{0-$\pi$} transitions it is necessary to understand their combined effect.
  From both analytical and numerical solutions of the Usadel equation we find that ripples can in fact produce a current response similar to that which is characteristic of a \mbox{0-$\pi$} transition.
  Hence, additional analysis is required in order to reveal the presence of a \mbox{0-$\pi$} transition caused by spin-splitting in graphene with ripples.
  We provide a closed form analytical expression for the critical current in the presence of exchange field and ripple effects as well as an expression for the scaling of critical current zeroes with junction parameters.
\end{abstract}
\maketitle
\section{Introduction}
\label{sec:introduction}
When the spinless superconducting order of a conventional superconductor comes in contact with a ferromagnet, it can adapt by creating spin-triplet Cooper pairs~\cite{eschrig_rpp_15}.
The synergy between superconductivity and ferromagnetism, two seemingly incompatible orders, is a topic of fundamental interest, but could also be of practical value.
One interesting consequence is that it allows for spinfull supercurrents.
The promise of low-dissipation spin transport has helped spawn the field of superconducting spintronics~\cite{linder2015}.

In the last decade, the possibility of creating Josephson junctions with graphene has attracted interest\cite{lee2018,Kumaravadivel2016,coskun2012,komatsu2012, heersche_nature_07}.
Superconductor-graphene-superconductor (SGS) junctions provide an arena for understanding the interplay between superconductivity and otherwise distinct physical phenomena, such as special relativity~\cite{lee2018} and the quantum Hall effect~\cite{Amet2016,liu2017}.
Here, we are interested in how the two-dimensional nature of monolayer graphene can be utilized to create an effective superconductor-ferromagnet-superconductor (SFS) junction with a tunable exchange field.
This is done by introducing a Zeeman splitting between the spin-bands in the graphene by use of a strong in-plane magnetic field.
This is possible because the two-dimensional nature of the graphene minimizes the magnetic depairing effect that would otherwise quench the superconducting correlations. By using electrodes with Ising-like superconductivity, like thin NbSe$_2$, one avoids destroying the superconducting state of the electrodes via the Pauli limitation.

The possibility of \emph{in situ} control of the exchange field could open new avenues for manipulations that take advantage of the combined effect of magnetic and superconducting order.
In addition to giving rise to the possibility of Cooper pairs with non-zero total spin~\cite{linder2015}, the presence of a magnetic field gives the Cooper pairs a non-zero total momentum, as first explained by Fulde, Ferrel, Larkin and Ovchinnikov~\cite{fulde1964,larkin1965}.
The total momentum of the Cooper pairs in the so-called FFLO-state is given by the strength of the exchange field, as this determines the displacement of the two Fermi surfaces corresponding to spin-up and spin-down electrons.
The non-zero momentum produces spatial variations in the superconducting order parameter~\cite{buzdin2005}. 

One consequence of the spatial variations, is that the ground state of a superconductor-ferromagnet-superconductor (SFS) Josephson junction can be one in which the phases of the order parameter in the two superconductors differs by $\pi$, which is known as a $\pi$-junction.
Such junctions could have an important role in the design of components for quantum computing~\cite{blatter2001,yamashita2005,khabipov2010}, superconducting computing~\cite{feofanov2010,ustinov2003} or as cryogenic memory~\cite{gingrich2016}.

Whether an SFS-junction is a $\pi$-junction or not depends on its length~\cite{oboznov2006,pugach2009} as well as the strength of the exchange field, which are typically fixed parameters. If, however, the exchange field could be tuned, this would allow for a controllable switching between the $0$-junction state and $\pi$-junction state.
Zeeman-effect-induced $\mbox{0-$\pi$}$ transitions have previously been observed in a Dirac semimetal with a $g$ factor on the order of $10^3$~\cite{li2019}.
The large $g$ factor allowed the $\mbox{0-$\pi$}$ transition to occur before the magnetic field extinguished the superconducting correlations.
Using a junction with a two-dimensional material, such as monolayer graphene, would allow for Zeeman driven $\mbox{0-$\pi$}$ transitions without the need for large $g$ factors, since such junctions can withstand much larger in-plane magnetic fields.

The prospect of a graphene Josephson junction being used as a tunable SFS-junctions is interesting, but it also demands a thorough investigation into how the supercurrent in an SGS-junction responds to the strong in-plane magnetic field necessary for an appreciable Zeeman splitting.
Even though monolayer graphene is two-dimensional, it will in general not be perfectly flat.
It will have a curvature that depends on the underlying substrate. If the graphene is placed on SiO$_2$, it will be rippled with peak-to-peak height difference of about $\SI{1}{\nano\meter}$ and typical feature size of $\SI{30}{\nano\meter}$~\cite{xue2011}.
Hence, the in-plane magnetic field will have a component orthogonal to the graphene surface, giving rise to an orbital effect.
This orthogonal component has been observed to suppress phase-coherent weak localization~\cite{lundeberg2010,zihlmann2020}.
Consequently, extra care must be taken when considering phase-coherent transport experiments relying on in-plane magnetic fields.
Here, we show that attention must also be paid to ripples when considering SGS-junction with tunable Zeeman splitting.

\begin{figure}[htpb]
  \centering
  \includegraphics[width=1.0\linewidth]{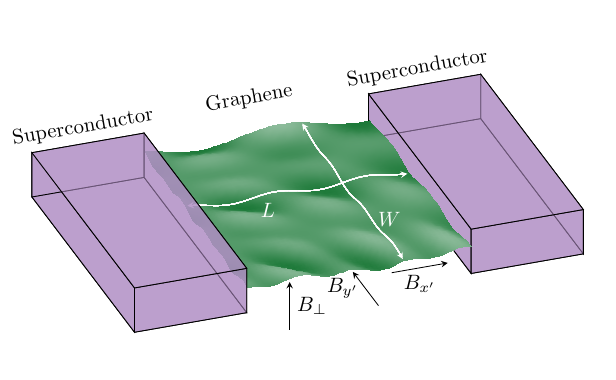}
  \caption{Sketch of superconductor-graphene-superconductor junction with rippled graphene and external magnetic field.
  The magnetic field has out-of-plane component $B_\perp$ and in-plane components $B_{x'}$ and $B_{y'}$.}%
  \label{fig:sketch}
\end{figure}

One property of a $\mbox{0-$\pi$}$ transition is that the current changes sign, giving zero net current exactly at the transition~\cite{oboznov2006}.
Characterization of an SGS-junction and the identification of a possible \mbox{0-$\pi$} transition will typically be done by measuring the current response to an applied magnetic field.
Therefore, it is important to know whether features in the critical current, such as the decay rate and zeros, can also be produced by an interference effect that arises from ripples in the graphene. 
Of particular interest is whether interference effects can give a vanishing critical current at magnetic field strengths that are comparable to the magnetic field necessary for a \mbox{0-$\pi$} transition.
The Zeeman energy necessary for the transition is typically smaller for diffusive systems~\cite{buzdin2005}, so diffusive systems, achievable for instance with SiO$_2$ as substrate~\cite{ke2016}, are most promising for tunable $\pi$-junctions. However, computing the supercurrent in a model with ripples in graphene is a challenging task due to the non-trivial geometry of the system.

A sketch of the system under consideration is shown in \cref{fig:sketch}.
In order to model this geometry, we add a spatially varying magnetic field to the equations governing diffusive SFS-junctions, where the exchange field in the ferromagnet comes from the Zeeman effect.
In order to fully model the disorderly ripples, one must be able to solve the equations with arbitrary magnetic field distributions.
We are able to do that numerically by using the finite element method.
Additionally, we extend the analytical result by \citet{bergeret2008} for superconductor-normal-superconductor (SNS) junctions with uniform magnetic fields to SFS-junctions with arbitrary magnetic field distributions and arbitrary exchange fields.

\section{Methodology}%
\label{sec:methodology}
The critical current, as well as other physical quantities such as the local density of states, can be calculated using the quasiclassical Keldysh Green's function formalism~\cite{usadel_prl_70, rammer_rmp_86}.
Previous studies of ballistic systems have investigated the interplay of the Zeeman and the orbital effect using an analytical propagator approach~\cite{hart2017,chen2018}.
The diffusive limit considered here, however, is more appropriately described by the Usadel equation, which previously has been done to successfully model experimental results for the supercurrent in SGS junctions~\cite{gueron_prb_16}.
We use natural units throughout, meaning that $c=\hbar =1$.

In thermal equilibrium it is sufficient to solve for the retarded Green's function, $\hat g$, which is normalized to $\hat g^2 = 1$ and solves the Usadel equation,
\begin{align}
  \label{eq:usadelGeneral}
  D\tilde \nabla\vdot \left(\hat g \tilde \nabla\hat g\right) + i\comm{\hat\Sigma}{\hat g} = 0,
\end{align}
provided that the Fermi wavelength and the elastic impurity scattering time is much shorter than all other relevant length scales, and the Fermi wavelength is much smaller than the scattering time. 
Here $D$ is the diffusion coefficient and the covariant derivative is 
$\tilde\nabla\hat g \equiv \nabla\hat g - ie\vb A \comm{\hat\rho_3}{\hat g}$,
where $\vb A$ is the vector potential and $e = -\abs{e}$ is the electron charge.
The self-energy is 
\begin{equation}
  \hat\Sigma = \begin{cases}
    (\varepsilon +i\delta) \hat\rho_3 + \mqty(\admat{\Delta, {-}\Delta, \cc{\Delta}, {-}\cc\Delta}) \quad \text{in the superconductors,}\\
    (\varepsilon +i\delta) \hat\rho_3 + \mqty(\dmat{\vb h \vdot \vb*{\sigma}, \vb h \vdot \cc{\vb*{\sigma}}}) \quad \text{in the ferromagnet,}
  \end{cases}
  \label{eq:selfEnergy}
\end{equation}
where $\varepsilon$ is the energy, $\delta^{-1}$ is the inelastic scattering time, $\hat\rho_3 = \diag(1,1,-1,-1)$, $\Delta$ is the superconducting gap parameter, $\vb h$ is the exchange field and $\vb* \sigma$ is the vector consisting of Pauli matrices.
Defining ``spin up'' and ``spin down'' parallel to the in-plane magnetic field gives $\vb h = h\vu z$.
We here disregard the effect of the very weak $\vb k$-dependent spin-orbit induced effective Zeeman-field caused by ripples in graphene, which is opposite in direction at the two inequivalent Dirac points of graphene~\cite{jeong_prb_11}.

To model rippled graphene, we will map the non-flat graphene sheet in the lab frame $x'y'z'$ shown in \cref{fig:sketch} with magnetic field components $B_{x'}, B_{y'}, B_\perp$ to a local coordinate system $xyz$ that follows the graphene surface in the sense that the $z$-axis is always orthogonal to the surface. The spatially constant magnetic field in the lab frame will then give rise to a spatially dependent magnetic field in the local coordinate system $xyz$.

Let $x$ and $y$ denote orthogonal coordinates on the graphene surface such that the superconductors are located at $x = \pm L/2$ and the interfaces with vacuum are at $y = \pm W/2$.
The curved geometry of graphene will in general enter into the Usadel equation~\eqref{eq:usadelGeneral} in three ways.
First, the divergence and gradient operators are altered by the curvature.
For a given ripple structure, these operators can be calculated from the resulting metric tensor.
However, since the curvature of rippled graphene typically is very small~\cite{xue2011}, the deviation from the Euclidean metric is negligible.
Consequently, we use $\nabla \equiv \vu x\pdv*{}{x} + \vu y\pdv*{}{y}$.
Second, the out-of-plane magnetic field component is modulated because it penetrates the graphene at varying angles.
That is, if $\vb B_\perp = B_\perp \vu z'$ is the out-of-plane magnetic field $\vu z'$ is the unit vector pointing out of the plane in the external coordinate system, or ``lab frame'', and $\vu z$ is the unit vector orthogonal to the graphene, then
\begin{equation}
  \vb B_\perp \vdot \vu z = B_\perp + \beta(x,y).
\end{equation}
Again, since the curvature is small, $\beta$ is negligible compared to $B_\perp$, so we can safely disregard this correction.
Third, and crucial for the effect considered here, the in-plane magnetic field has a component orthogonal to the surface.
In the lab frame, the magnetic field can be written $\vb B = \transpose{(B_{x'}, B_{y'}, B_\perp)}$.
In the coordinate system of the curved graphene, this gives rise to a magnetic field with a $z$-component equal to
\begin{multline}
  \vb B \vdot \vu z = -B_{x'}\sin\left[\arctan(\pdv{\eta}{x'})\right]
        -B_{y'}\sin\left[\arctan(\pdv{\eta}{y'})\right] \\ 
        + B_\perp + \beta
    \approx -B_{x'} \pdv{\eta}{x}-B_{y'} \pdv{\eta}{y}+ B_\perp,
    \label{eq:magFieldLocal}
\end{multline}
where $\eta$ is the height distribution of the graphene, $x'$ and $y'$ are the coordinates in the lab frame and we have used the assumption that $\pdv*{\eta}{x} \approx \pdv*{\eta}{x'} \ll 1$ and $\pdv*{\eta}{y}\approx \pdv*{\eta}{y'} \ll 1$. 

In order to capture the magnetic field given by \cref{eq:magFieldLocal}, we use the vector potential
 \begin{equation}
   \vb A = \left(B_{y'} \eta + B_{x'} \int_0^y \pdv{\eta}{x}\dd{\tilde y} - B_\perp y\right)\vu x.
   \label{eq:vecPot}
 \end{equation}
 Choosing the vector potential parallel to the $x$-axis allows us to set
 $\vb A = \vb 0$ in the superconductors, which means that the ground states in the superconductors have constant phases.
 In the following we denote the superconducting phase in the left ($x < -L/2$) and right ($x > L/2$) superconductors by $\phi_l$ and $\phi_r$, respectively. Having established the form of the vector potential $\vb A$ [\cref{eq:vecPot}] in the local coordinate system where graphene is flat, we will from now on omit the prime on $B_{x'}$ and $B_{y'}$ for brevity of notation.

The Usadel equation~\eqref{eq:usadelGeneral} is not valid across boundaries of different materials since the associated length scales are not negligible compared to the Fermi wavelength.
Instead, the Green's function in the different materials must be connected through a boundary condition.
For low-transparency tunneling interfaces, one may use the
Kupriyanov-Lukichev boundary condition~\cite{kuprianov1988},
\begin{align}
  \label{eq:KL}
  \zeta_i L_i \vu{n} \vdot \left(\hat g_i\tilde\nabla\hat g_i\right) = \frac 1 2 \comm{\hat g_i}{\hat g_j},
\end{align}
where the subscripts $i$ and $j$ denote the different sides of the interface, $\vu n$ is a normal unit vector pointing out of region $i$, $L_i$ is the length of region $i$ in the $\vu n$-direction and $\zeta_i$ is the ratio of the normal-state conductance of region $i$ to the interface conductance.
\Cref{eq:KL} is used along the interface between the superconductors and graphene at $x=-L/2$ and $x=L/2$.
Along the boundaries with vacuum at $y = \pm W/2$, the boundary condition is $\vu n \vdot \tilde\nabla \hat g = 0$.

It has been shown that one may use the bulk solution in the superconductors,
\begin{equation}
\hat g_\textsc{bcs} = \frac{\hat \Sigma}{\sqrt{(\varepsilon + i\delta)^2 - \abs{\Delta}^2}},
\label{eq:scBulk}
\end{equation}
when the interface conductance is much smaller than the normal-state conductance of length $\xi$ of the superconductor~\cite{fyhn2019}, where 
\begin{equation}
  \xi = \sqrt{\frac{D}{\Delta}}
\end{equation}
is the coherence length.
The square root in \cref{eq:scBulk} must be chosen such that it has a positive imaginary part.

Having found the Green's function, the electrical current density can be calculated from~\cite{Belzig1999}
\begin{equation}
  \vb{j} = \frac{N_0eD}{4}\int_{-\infty}^{\infty}\Trace\left(\hat\rho_3\hat g\tilde\nabla \hat g - \hat g^\dagger\tilde\nabla\hat g^\dagger\hat\rho_3\right)\tanh\left(\frac{\beta\varepsilon}{2}\right)\dd \varepsilon,
  \label{eq:currDen}
\end{equation}
where $N_0$ is the normal density of states and $\beta$ is inverse temperature.
Finally, \cref{eq:currDen} allows for calculation of the critical current, given by
\begin{equation}
  I_c = \max_{\phi_r - \phi_l} \int_{-W/2}^{W/2} \vu x \vdot \vb j(x, y) \dd{y},
  \label{eq:critCurr}
\end{equation}
where the choice of $x$ is arbitrary.

In order to solve \cref{eq:usadelGeneral} numerically, we use the Ricatti parametrisation,
\begin{equation}
  \hat g = \mqty(\dmat{N, -\tilde N})\mqty(1+\gamma\tilde\gamma & 2\gamma \\ 2\tilde\gamma & 1+\tilde\gamma\gamma),
  \label{eq:ricatti}
\end{equation}
where $N = (1-\gamma\tilde\gamma)^{-1}$ and tilde conjugation is defined as $\tilde\gamma(\varepsilon) = \cc\gamma(-\varepsilon)$.
This respects the normalization and underlying symmetries of $\hat g$.
The resulting equations for the $2\times 2$ matrices $\gamma$ and $\tilde\gamma$ are discretized by the finite element method~\cite{Amundsen2016} with quadratic elements, and Gauss-Legendre quadrature rules of fourth order is used to integrate over the elements.
The resulting nonlinear set of algebraic equations are solved by the Newton-Raphson method~\cite{burden2011}, where the Jacobian is determined by forward-mode automatic differentiation~\cite{RevelsLubinPapamarkou2016}.

\section{Results and Discussion}%
\label{sec:results}
In order to linearize the Usadel equation, we write
\begin{equation}
  \hat g = \hat g_0 + \hat f,
\end{equation}
where $\hat g_0$ is the bulk solution.
In a ferromagnet, the self-energy $\hat\Sigma$, given by \cref{eq:selfEnergy}, is diagonal.
Hence, the bulk equation $\comm{\hat\Sigma}{\hat g_0} = 0$ is solved by any diagonal matrix satisfying $\hat g_0^2 = 1$.
In order to find the correct solution one must solve the full Gor'kov equation.
The result is that $\hat g_0 = \hat \rho_3$.
If we assume that the proximity effect is weak, we can keep only linear terms in $\hat f$, yielding
\begin{equation}
  D\hat\rho_3\tilde\nabla^2 \hat f + i\comm{\Sigma}{\hat f} = 0,
\end{equation}

In order for the normalization $\hat g^2 = 1$ to hold to linear order in $\hat f$, we need $\acomm{\hat f}{\hat\rho_3} = 0$.
This implies that
\begin{equation}
  \hat f = \mqty(0 & f \\ -\tilde f & 0)
  \label{eq:fshape}
\end{equation}
and
\begin{equation}
  \tilde\nabla \hat f = \nabla\hat f - 2ie\vb A \hat\rho_3\hat f.
\end{equation}

In the weak proximity effect regime, the boundary conditions read
\begin{equation}
  \zeta L \hat\rho_3 \vu{n} \vdot \tilde\nabla\hat f = \frac 1 2 \comm{\hat \rho_3 + \hat f}{\hat g_s + \hat f_s}
  = \hat\rho_3 \hat f_s  - \hat g_s \hat f,
  \label{eq:weakBC}
\end{equation}
where $\hat g_s$ is the part of $\hat g_\textsc{bcs}$ proportional to $\hat\rho_3$ and $\hat f_s$ is the remaining part proportional to $\antidiag(\Delta, -\Delta, \cc\Delta, -\cc\Delta)$.
Additionally, we must have $\comm{\hat f}{\hat f_s} = 0$, since this term would be block diagonal.

Since $\hat\Sigma$ is diagonal in the ferromagnet, the different components of $\hat f$ decouple.
Only the elements which are nonzero in $\hat f_s$ will have a constant term in the boundary conditions. 
The remaining elements must be zero.
Hence, $\hat f$ must be antidiagonal, just like $\hat f_s$.
This in turn implies that $\comm{\hat\Sigma}{\hat f} = 2\hat\Sigma\hat f$.

In order to solve the Usadel equation for arbitrary magnetic fields, we first define
\begin{equation}
  \hat u = \exp(-2ie\hat\rho_3\int_{-L/2}^x A(x', y) \dd{x'})\hat f
\end{equation}
With this, the Usadel equation can be written
\begin{multline}
  D\hat\rho_3\nabla^2\hat u + 2i\hat \Sigma\hat u - 2Die\hat\rho_3\hat u\int_{-L/2}^x\pdv{B}{y} \dd{x'} \\
  -4D e^2 \hat u \left(\int_{-L/2}^x B \dd{x'}\right)^2
  - 4Die \pdv{\hat u}{y}\int_{-L/2}^x B \dd{x'} = 0,
  \label{eq:usadelWithMag}
\end{multline}
where we have defined $B \equiv B_z = -\pdv*{A}{y}$ as the magnetic field component orthogonal to the graphene.
We can neglect the terms involving $B$ by assuming that the magnetic field is sufficiently weak. 
That is, for all $x\in (-L/2, L/2)$,
\begin{subequations}
\label{eq:magAssum}
\begin{align}
  \int_{-L/2}^x B \dd{x'} \ll \Phi_0 \sqrt{\frac{\delta}{D}}, \\
  \intertext{and}
  \int_{-L/2}^x\pdv{B}{y} \dd{x'} \ll \frac{\Phi_0 \delta}{D},
\end{align}
\end{subequations}
where $\Phi_0 = \pi/e$ is the magnetic flux quantum.

The boundary conditions for $\hat u$ at $y = \pm W/2$ is
\begin{equation}
  \eval{\pdv{\hat u}{y}}_{y = \pm W/2} = -i\hat u\int_{-L/2}^x B \dd{x'}.
  \label{eq:BCuVacuum}
\end{equation}
\Cref{eq:weakBC,eq:usadelWithMag,eq:BCuVacuum} can be solved exactly when $B = 0$ by assuming $\pdv*[2]{\hat u}{y} = 0$.
For $B$ satisfying \cref{eq:magAssum} we can find an approximate solution by neglecting the term $\pdv*[2]{\hat u}{y}$ in \cref{eq:usadelWithMag}.

With these approximations, the Usadel equation becomes an ordinary differential equation,
\begin{equation}
  \pdv[2]{\hat u}{x} + \frac{2i\hat\rho_3\hat\Sigma}{D}\hat u = 0,
\end{equation}
with solution
\begin{equation}
  % \hat u = \exp(\sqrt{\frac{2i\hat\rho_3\hat\Sigma}{D}}x)\hat A + \exp(-\sqrt{\frac{2i\hat\rho_3\hat\Sigma}{D}}x)\hat B
  \hat u = \me{\hat k x}\hat A + \me{-\hat k x}\hat B,
\end{equation}
for some coefficients $\hat A$ and $\hat B$.
Here 
\begin{equation}
 \hat k = \sqrt{-\frac{2i\hat\rho_3\hat\Sigma}{D}},
\end{equation}
which, since $\hat\rho_3\hat\Sigma$ is diagonal, $\hat k$ can be obtained simply by taking the elementwise square root.
To determine $\hat A$ and $\hat B$ one must use \cref{eq:weakBC}. The solution is
\begin{multline}
  % \hat u = \frac{\hat p(x) + \me{i\hat\rho_3\theta}\hat p(-x)}{(\zeta L\hat k + \hat\rho_3\hat g_s)^2\me{\hat k L} - (\zeta L\hat k + \hat\rho_3\hat g_s)^2\me{-\hat k L}}\hat fs,
  \hat u = \left[(\zeta L\hat k + \hat\rho_3\hat g_s)^2\me{\hat k L} - (\zeta L\hat k - \hat\rho_3\hat g_s)^2\me{-\hat k L}\right]^{-1} \\
  \times \left[\hat p(x) + \me{i\hat\rho_3\theta}\hat p(-x)\right]\hat f_s,
  \label{eq:solu}
\end{multline}
where
\begin{equation}
  \theta = \phi_r - \phi_l - 2e\int_{-L/2}^{L/2} A\left(x, y\right)\dd{x}
  \label{eq:theta}
\end{equation}
and
\begin{equation}
  \hat p(x) = (\zeta L\hat k + \hat\rho_3 \hat g_s)\me{\hat k (x - L/2)}  + (\zeta L \hat k - \hat \rho_3 \hat g_s)\me{-\hat k (x-L/2)}.
\end{equation}
Note that the boundary condition for the interfaces with vacuum, \cref{eq:BCuVacuum}, is only approximately satisfied.

To find the current from \cref{eq:solu} we can use that the $x$-component of the current, as given by \cref{eq:currDen} can be written
\begin{equation}
  \vb{j}\vdot\vu x = \frac{N_0eD}{4}\int_{-\infty}^{\infty}\Trace\left(\hat\rho_3\hat u\pdv{\hat u}{x} - \hat u^\dagger\pdv{\hat u^\dagger}{x}\hat\rho_3\right)\tanh\left(\frac{\beta\varepsilon}{2}\right)\dd{\varepsilon}
  \label{eq:currU}
\end{equation}
To simplify this expression, note that \cref{eq:solu} can be written
\begin{equation} 
  \hat u = \mqty(\admat{d(h)\me{i\phi_l}, {-}d({-}h)\me{i\phi_l},{-}\tilde d(h)\me{-i\phi_l}, \tilde d({-}h)\me{-i\phi_l}})
\end{equation}
where
\begin{equation}
  d(h) = \frac{[p(x) + \me{i\theta}p(-x)]\abs{\Delta}/\sqrt{(\varepsilon + i\delta)^2 - \abs{\Delta}^2}}{\left[(\zeta L k + g_s)^2\me{k L} - (\zeta L k - g_s)^2\me{-k L}\right]}
  \label{eq:d}
\end{equation}
with $g_s = (\varepsilon + i\delta)/\sqrt{(\varepsilon + i\delta)^2 - \abs{\Delta}^2}$, the square roots are those which have positive imaginary parts, $k = \sqrt{-2i(\varepsilon + h + i\delta)/D}$ and
\begin{equation}
  p(x) = (\zeta Lk + g_s)\me{k (x - L/2)}  + (\zeta L k - g_s)\me{-k (x-L/2)}.
\end{equation}
Inserting \cref{eq:d} into \cref{eq:currU} gives
\begin{multline}
  \vb j \vdot \vu x = N_0 eD \int_{-\infty}^{\infty}\Re\left[d(h)\pdv{\tilde d({-}h)}{x} + d({-h})\pdv{\tilde d(h)}{x}\right] \\
  \times \tanh\left(\frac{\beta\varepsilon}{2}\right)\dd \varepsilon.
  \label{eq:currd}
\end{multline}

By evaluating $\vb j$ at $x = 0$ we can factorize out the dependence on the vector potential, since
\begin{equation}
  \eval{\pdv{\tilde d(h)}{x}}_{x=0} = -\eval{\frac{2i\sin\theta}{1+\me{i\theta}}\times\pdv{\ln p}{x} d({-}h)}_{x=0}.
\end{equation}
Inserting this into \cref{eq:currd} and integrating the current density over $y$ to obtain the total current finally gives
\begin{widetext}
  \begin{align}
  I = 2N_0eD \int_{-W/2}^{W/2}\sin\left(\phi_r - \phi_l - 2e\int_{-L/2}^{L/2} A\left(x, y\right)\dd{x}\right)\dd{y}
  \times \int_{-\infty}^{\infty}\Im\left[\kappa_+ + \kappa_-\right]\tanh\left(\frac{\beta\varepsilon}{2}\right) \dd \varepsilon,
  \label{eq:currSol}
  \end{align}
% \end{widetext}
% \begin{multline}
%   I = 2N_0eD \int_{-W/2}^{W/2}\sin\left(\phi_r - \phi_l + 2e\int_{-L/2}^{L/2} A\left(x, y\right)\dd{x}\right)\dd{y} \\
%   \times \int_{-\infty}^{\infty}\Im\left[\kappa_+ + \kappa_-\right]\tanh\left(\frac{\beta\varepsilon}{2}\right) \dd \varepsilon,
%   \label{eq:currSol}
% \end{multline}
where
% \begin{widetext}
\begin{align}
  \kappa_\pm = \frac{4\abs{\Delta}^2k_\pm}{(\varepsilon + i\delta)^2 - \abs{\Delta}^2}
  \times \frac{\left[\zeta Lk_\pm\cosh(k_\pm L/2) - g_s\sinh(k_\pm L/2)\right]\times\left[g_s\cosh(k_\pm L/2) -  \zeta Lk_\pm\sinh(k_\pm L/2)\right]}{\left[(\zeta L k_\pm + g_s)^2\me{k_\pm L} - (\zeta L k_\pm - g_s)^2\me{-k_\pm L}\right]^2},
\end{align}
\end{widetext}
with $k_\pm = \sqrt{-2i(\varepsilon \pm h + i\delta)/D}$.

\begin{figure}[htpb]
  \centering
  \includegraphics[width=1.0\linewidth]{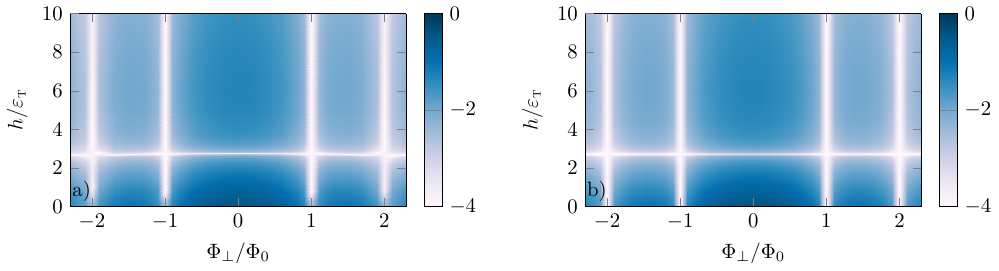}
  \caption{Color plot of $\log_{10}(I_c/I_0)$, where $I_c$ is the critical current and $I_0 = N_0 e WD^2/L^3$ for various exchange fields $h$ and out-of-plane magnetic fluxes.
  Figure a) shows the solution found numerically from the full Usadel equation, and b) shows the analytical solution found using \cref{eq:currSol}.
Here, $\thouless = D/L^2$, $W = 10L$, $L = 6\xi$, $\delta/\Delta = 0.01$ and $B_x=B_y= 0$.}
    \label{fig:uniform}
\end{figure}

\Cref{eq:currSol} is our main analytical result and allows for evaluation of the current at arbitrary exchange field strengths and magnetic field distributions.
Of particular interest is the fact that the contributions from the vector potential and exchange field decouple.
One consequence of this is that a constant magnetic field gives rise to a Fraunhofer pattern in the current regardless of the strength of the exchange field, as long as the magnetic field is weak enough.
This can be seen from the fact that for a constant magnetic field $eA = -\pi \Phi_\perp y/\Phi_0 WL$, where $\Phi_\perp$ is the magnetic flux from the perpendicular field, so
\begin{multline}
  I \propto \int_{-W/2}^{W/2}\sin(\phi_r -\phi_l + 2\pi \frac{\Phi_\perp y}{\Phi_0W})\dd{y} \\ 
  = W\sin(\phi_r - \phi_l)\frac{\sin(\pi \Phi_\perp/\Phi_0)}{\pi \Phi_\perp/\Phi_0}.
\end{multline}
\Cref{fig:uniform} shows the critical current found analytically using \cref{eq:currSol} for the case of no ripples, compared to the critical current obtained numerically from the full nonlinear Usadel equation.
In addition to showing the agreement between the full solution and the analytical approximation, \cref{fig:uniform} also shows that there is an exchange-driven \mbox{0-$\pi$} transition at $h \approx 2\thouless$, where
\begin{equation}
  \thouless = \frac{D}{L^2}
\end{equation}
is the Thouless energy.

Since the contribution from the exchange field is independent of the vector potential, we can focus on how the magnetic field alters the critical current.
With the vector potential given by \cref{eq:vecPot}, we get
\begin{multline}
  \int_{L/2}^{L/2}A(x,y)\dd{x} = B_y L\bar \eta(y) -\Phi_\perp \frac{y}{W} \\ 
  + B_x \int_{0}^y\left[\eta(L/2, \tilde y) - \eta(-L/2, \tilde y)\right]\dd{\tilde y},  
  \label{eq:vecPotIntegral}
\end{multline}
where
\begin{equation}
  \bar \eta(y) = \frac 1 L \int_{-L/2}^{L/2}\eta(x, y)\dd{x}
  \label{eq:longAvHeight}
\end{equation}
is the longitudinally averaged height.
From \cref{eq:vecPotIntegral} it can be observed that the contribution proportional to $B_x$ is small for variations that are fast in the $y$-direction, since the integrand will oscillate rapidly, and small for very slow variations, which will contribute little to $\eta(L/2, y) - \eta(-L/2, y)$.
Similarly, variations that are fast in the $x$-direction will contribute little to the term proportional to $B_y$.
Otherwise, the contributions from the terms proportional to $B_y$ and $B_x$ is similar, so we set $B_x = 0$ in the following.
We also set $\beta = 1000/\Delta$, corresponding to $T/T_c \approx 1.8\times10^{-3}$, where $T_c$ is the critical temperature. 

From \cref{eq:currSol,eq:vecPotIntegral} we can find how big the height variations must be in order to possibly cause a vanishing critical current at $\Phi_\perp = 0$.
In order for the critical current to vanish, the argument of the sine function in \cref{eq:currSol} must have variations of at least $\pi/2$.
Otherwise, the phase difference, $\phi_r - \phi_l$, can be chosen such that the integrand is of one sign.
This means that in order for there to be a root in the critical current at $\Phi_\perp = 0$, the in-plane magnetic field must be at least
\begin{equation}
  B_y = \frac{\Phi_0}{4L(\max\bar\eta - \min\bar\eta)},
  \label{eq:magReq}
\end{equation}
assuming $B_x = 0$.

In order to apply \cref{eq:currSol} to the case of rippled graphene with vector potential given by \cref{eq:vecPot}, we need a model of the height distribution $\eta$ of the ripples.
From \cref{eq:vecPotIntegral} we find that it is reasonable to categorize ripples into short ripples and long ripples, depending on whether the wavelength is shorter or longer than $2L$.
Ripples with wavelength shorter than $2L$ will have a smaller contribution to $\bar\eta$ in \cref{eq:vecPotIntegral} since the integrand oscillates between positive and negative values.
For this reason, short ripples will contribute less to interference effects than long ripples with the same amplitude.
On the other hand, faster variations in the $y$-direction gives a larger magnetic field component perpendicular to the graphene surface, and therefore a larger depairing effect.
Short ripples are therefore expected to lead to larger deviations from the analytical approximation given by \cref{eq:currSol}.
In particular, they are expected to cause a faster decay, which, as we will see, is also what happens.

In general, the height distribution will be a superposition of long and short ripples.
We look first at only long ripples, then at only short ripples and finally at the combination of both short and long ripples.
In order to simplify the presentation and analysis, we present solutions for height distributions that can be written as product of cosines.
We obtain qualitatively similar result for more realistic, randomized height distributions.

\begin{figure}[htpb]
  \centering
  \includegraphics[width=1.0\linewidth]{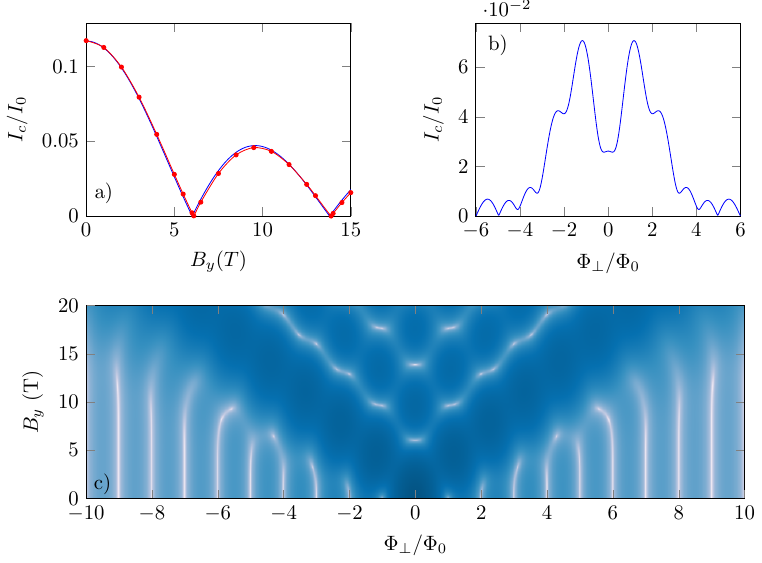}
  \caption{Critical current $I_c$ for various in-plane magnetic field strengths $B_y$ and out-of-plane magnetic fluxes $\Phi_\perp$ for the height distribution given by \cref{eq:cosHeight} with $n=1$ and $m=2$.
  Here, $I_0 = N_0 e WD^2/L^3$, $W = 10L$, $L = 10\xi = \SI{400}{\nano\meter}$, $\eta_0 = \SI{1}{\nano\meter}$, $\delta/\Delta = 0.02$ and $B_x = 0$.
  Figure a) shows the analytical solution of $I_c$ against $B_y$ for $\Phi_\perp = 0$ (blue line), compared to the numerical solution (red dots). b) shows analytical solution of $I_c$ against $\Phi_\perp$ for $B_y = \SI{5}{\tesla}$ and c) shows a logarithmically scaled color plot of analytical $I_c$ where white means zero current and deep blue corresponds to large current.}%
  \label{fig:cos1_2}
\end{figure}
\begin{figure}[htpb]
  \centering
  \includegraphics[width=1.0\linewidth]{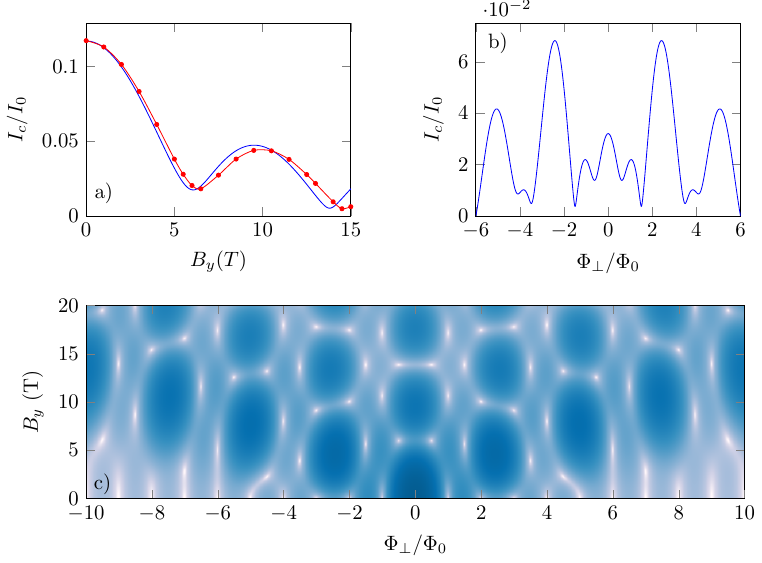}
  \caption{Critical current $I_c$ for various in-plane magnetic field strengths $B_y$ and out-of-plane magnetic fluxes $\Phi_\perp$ for the height distribution given by \cref{eq:cosHeight} with $n=1$ and $m=5$.
  Here, $I_0 = N_0 e WD^2/L^3$, $W = 10L$, $L = 10\xi = \SI{400}{\nano\meter}$, $\eta_0 = \SI{1}{\nano\meter}$, $\delta/\Delta = 0.02$ and $B_x = 0$.
  Figure a) shows the analytical solution of $I_c$ against $B_y$ for $\Phi_\perp = 0$ (blue line), compared to the numerical solution (red dots). b) shows analytical solution of $I_c$ against $\Phi_\perp$ for $B_y = \SI{5}{\tesla}$ and c) shows a logarithmically scaled color plot of analytical $I_c$ where white means zero current and deep blue corresponds to large current.}%
  \label{fig:cos1_5}
\end{figure}
\begin{figure}[htpb]
  \centering
  \includegraphics[width=1.0\linewidth]{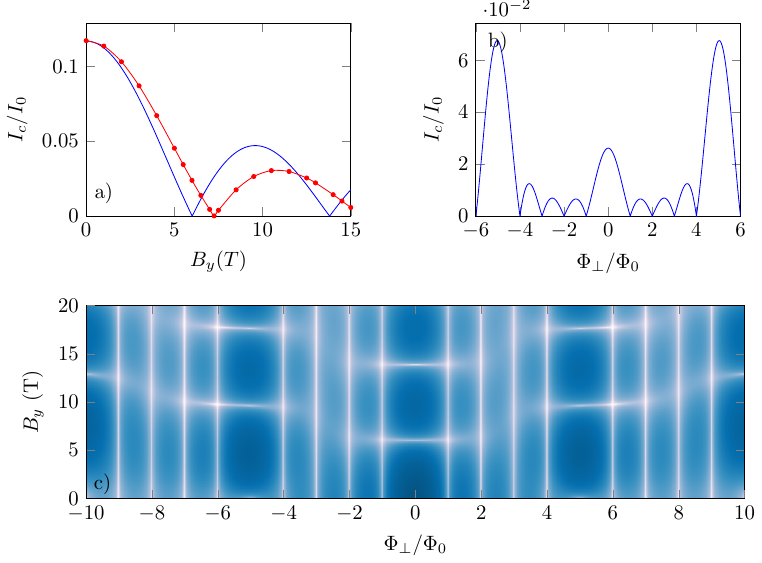}
  \caption{Critical current $I_c$ for various in-plane magnetic field strengths $B_y$ and out-of-plane magnetic fluxes $\Phi_\perp$ for the height distribution given by \cref{eq:cosHeight} with $n=1$ and $m=10$.
  Here, $I_0 = N_0 e WD^2/L^3$, $W = 10L$, $L = 10\xi = \SI{400}{\nano\meter}$, $\eta_0 = \SI{1}{\nano\meter}$, $\delta/\Delta = 0.02$ and $B_x = 0$.
  Figure a) shows the analytical solution of $I_c$ against $B_y$ for $\Phi_\perp = 0$ (blue line), compared to the numerical solution (red dots). b) shows analytical solution of $I_c$ against $\Phi_\perp$ for $B_y = \SI{5}{\tesla}$ and c) shows a logarithmically scaled color plot of analytical $I_c$ where white means zero current and deep blue corresponds to large current.}%
  \label{fig:cos1_10}
\end{figure}

To model $n$ uniform ripples in the $x$-direction and $m$ uniform ripples in the $y$-direction, we use the height distribution
\begin{equation}
  \eta(x,y) = \frac{\eta_0}{2} \cos(n\pi\frac x L)\cos(m\pi \frac y W),
  \label{eq:cosHeight}
\end{equation}
where $\eta_0$ is the peak-to-peak height difference.
\Cref{fig:cos1_2,fig:cos1_5,fig:cos1_10} show the critical current, given by \cref{eq:critCurr},
for the height distribution in \cref{eq:cosHeight} with $n=1$ and $m=2$, $m=5$ and $m=10$, respectively.
The exchange field, $h$, is set to zero in order to isolate the orbital effect, such that it can be observed whether the orbital effect alone is sufficient to produce roots in the critical current.
Physically, the situation with negligible exchange field would be the case if the Thouless energy, $\thouless$, is much larger than the Zeeman splitting $\mu_B B_y$, where $\mu_B$ is the Bohr magneton.

It can be seen from \cref{fig:cos1_2,fig:cos1_5,fig:cos1_10} that a height variation of only $\SI{1}{\nano\meter}$ is sufficient to produce oscillations in the critical current.
In particular, the critical current is zero for $\Phi_\perp = 0$ and finite $B_y$ when $m=2$ and $m=10$.
Both zeros satisfies \cref{eq:magReq}, which is $B_y = \SI{2}{\tesla}$ in this case.
We can conclude from this that a zero in the critical current of an SGS junction is not sufficient to identify an exchange-driven \mbox{0-$\pi$} transition, since zeros can also be produced from the ripples.

Since the junction widths are equal in \cref{fig:cos1_2,fig:cos1_5,fig:cos1_10}, larger $m$ means a larger orthogonal component from the in-plane magnet field.
Consequently, the magnetic field strengths for which the weak field assumption, \cref{eq:magAssum}, remains valid is reduced when $m$ increases.
This is reflected in the correspondence between the numerical simulations and the analytical predictions.

The color plots in \cref{fig:cos1_2,fig:cos1_5,fig:cos1_10}~c) show that the critical current has especially large maxima at $\Phi_\perp/\Phi_0 = km/2$ for integer $k$.
This is also reflected in \cref{fig:cos1_2,fig:cos1_5,fig:cos1_10}~b), which show that the lobe structure has strong maxima at $\Phi_\perp/\Phi_0 = m/2$.
To understand why, note that the orbital part of the current can be written as a Fourier transform. That is,
\begin{multline}
  I = C(h)\int_{-W/2}^{W/2}\sin(\phi_l -\phi_r + 2\pi\frac{B_y L \bar\eta}{\Phi_0} - 2\pi \frac{\Phi_\perp y}{\Phi_0W})\dd{y} \\
  = C(h)\Im\left\{\me{i(\phi_l -\phi_r)}\int_{-W/2}^{W/2}\exp[2\pi i\left(\frac{B_y L \bar\eta}{\Phi_0} - \frac{\Phi_\perp y}{\Phi_0W}\right)]\dd y\right\},
\end{multline}
where $C$ is a function of the exchange field.
Hence, the critical current can be written as
\begin{equation}
  I_c = C(h)\abs{\mathcal{F}\left[\rect(y/W)\me{2\pi iB_yL\bar\eta/\Phi_0}\right]\left(\frac{2\pi\Phi_\perp}{W\Phi_0}\right)},
  \label{eq:fourierVers}
\end{equation}
where $\rect$ is the rectangular function and $\mathcal F$ means Fourier transform.
Accordingly, a Fourier analysis of the current response to out-of-plane magnetic fields can uncover properties of the ripple structure.
In this case, $\bar\eta$ is a cosine with wavenumber $m\pi/W$, so it is reasonable that the Fourier transform peaks at $2\pi\Phi_\perp/W\Phi_0 = km\pi/W$ with strengths that depends on $B_y$.
For a more general ripple structure, mapping out the current response to both in-plane and out-of-plane one could uncover information about the slow height variations that can be difficult to detect with surface probe techniques.

Obtaining $\bar\eta$ from measurements of the critical current is complicated by the loss of the phase information of the Fourier transform in \cref{eq:fourierVers}.
One possible resolution is to obtain the current-phase relation, as has been done for ballistic graphene by use of a SQUID~\cite{nanda2017}.
In this case, one could take advantage of the fact that the current for a given in-plane magnetic field is proportional to $\sin(\phi_l-\phi_r)\Re(F) + \cos(\phi_l-\phi_r)\Im(F)$, where $F$ is the Fourier transform in \cref{eq:fourierVers}.
It would then in theory be straightforward to find the phase of $F$ and compute its Fourier inverse.
If the only available data is the critical current, one could look at the values of $\Phi_\perp$ that gives enhanced current upon application of in-plane magnetic field to extract the most prominent Fourier components of $\bar\eta$ or use the variational method to approximate the function $\bar\eta$ that solves \cref{eq:fourierVers}.
If one can also manipulate the direction of the in-plane magnetic field, one could combine the knowledge of $\bar\eta$ with the corresponding function relevant for when the field is in the $x$-direction to get even more insight into the full ripple profile $\eta$.

\Cref{fig:cos1_2,fig:cos1_5,fig:cos1_10}~a) also give clues to how the full solution of the Usadel equation deviates from \cref{eq:currSol} when the magnetic field is strong.
Two things seem to happen when the flux density is strong.
First, compared to the analytical solution, the full solution decays more rapidly as $B_y$ increases.
That the critical current decays faster than the analytical solution predicts is unsurprising, since we neglected the depairing effect of the magnetic field in our derivation of \cref{eq:currSol}.

Second, the functional dependence on $B_y$ is slower in the numerical case, in the sense that roots and extremal values in the critical current are skewed towards larger values of $B_y$.
A plausible explanation for this phenomenon is that the full solution varies more slowly in the $y$-direction compared to the analytical approximation in \cref{eq:solu}.
The analytical solution has $\partial \hat f/\partial y \propto B$, but it neglects the boundary condition that demands $\partial \hat f/\partial y = 0$ when $y = \pm W/2$.
Hence, it is possible that the analytical solution overestimates the variation of $f$ with respect to $y$, at least close to $y = \pm W/2$.
The roots in the critical current occur because the magnetic field creates mutually cancelling oscillations in the current density as a function of $y$.
If the analytical approximation overestimates how fast these oscillations occur, it will underestimate the magnetic field required to give $I_c = 0$.
The faster decay, but slower variation is also exactly what happens in the case of uniform magnetic fields, as can be seen from fig. 3 in Ref.~\cite{bergeret2008}.

\begin{figure}[htpb]
  \centering
  \includegraphics[width=1.0\linewidth]{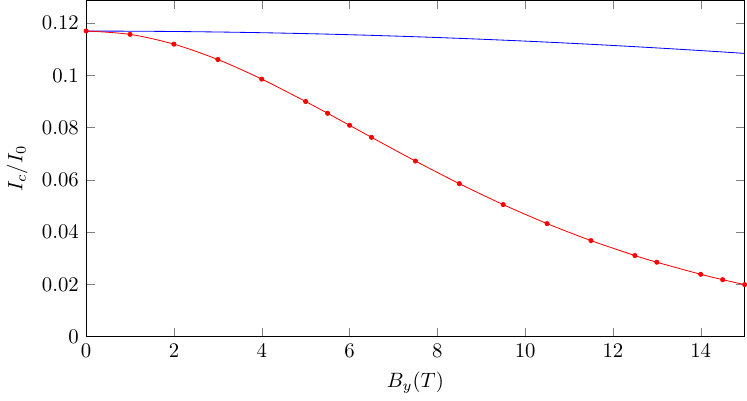}
  \caption{Critical current $I_c$ for $B_\perp = 0$ and various in-plane magnetic field strengths $B_y$ with the height distribution given by \cref{eq:cosHeight} with $n=11$ and $m=10$.
  Here, $I_0 = N_0 e WD^2/L^3$, $W = L = 10\xi = \SI{400}{\nano\meter}$, $\eta_0 = \SI{1}{\nano\meter}$, $\delta/\Delta = 0.02$ and $B_x = 0$.
  The blue line shows the analytical solution, as found by \cref{eq:currSol}, and the red line with dots shows the numerical solution found by solving the full Usadel equation.}%
  \label{fig:cos11_10}
\end{figure}

Moving on to short ripples, \cref{fig:cos11_10} shows the critical current for the distribution given by \cref{eq:cosHeight} with $n=11$, $m=10$ and $\eta_0 = \SI{1}{\nano\meter}$. With $L = \SI{400}{\nano\meter}$, this corresponds to a  ripple length of $\SI{40}{\nano\meter}$, which is comparable to the short ripples observed in graphene on SiO$_2$~\cite{xue2011}.
The exchange field is again set to 0.
What matters for the analytical approximation is the longitudinally averaged height, $\bar\eta$, which in this case is small because of the rapid oscillations.
Hence, the analytical approximation predicts very little change in the critical current at $\Phi_\perp = 0$.
On the other hand, the orbital depairing effect is quite large because of the short ripples, which is reflected in the decay of the critical current observed in the numerical solution of the full Usadel equations.

\begin{figure}[htpb]
  \centering
  \includegraphics[width=1.0\linewidth]{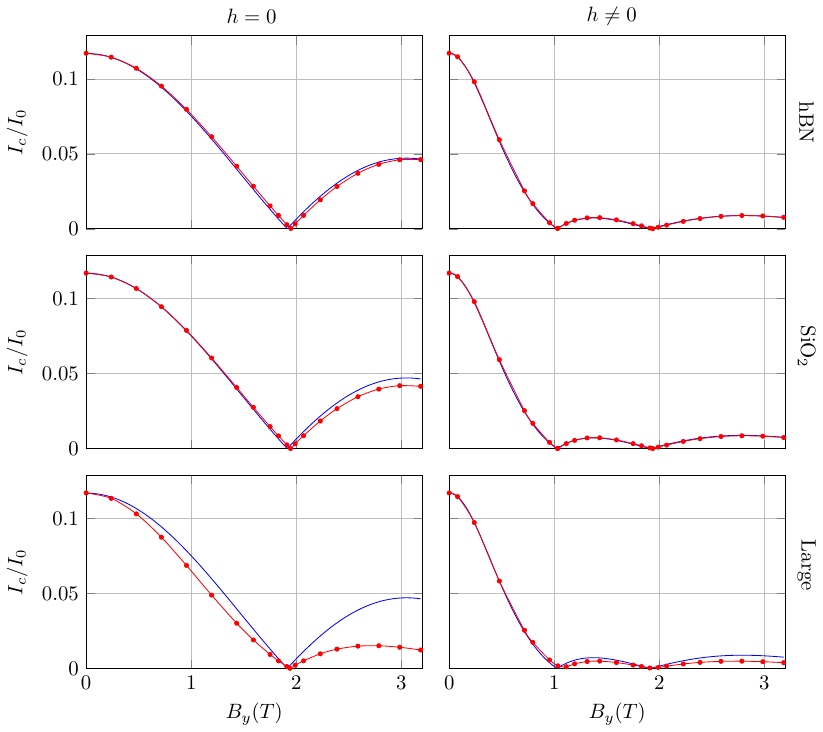}
  \caption{Critical current $I_c$ for $B_\perp = 0$ and various in-plane magnetic field strengths $B_y$ with the height distribution given by \cref{eq:combined} with $(A, N) = (\SI{0.1}{\nano\meter}, 4)$, $(\SI{0.5}{\nano\meter}, 10)$ and $(\SI{2}{\nano\meter}, 4)$, respectively from top to bottom.
    The exchange field is $h=0$ in the left panels and $h=\mu_B B_y$ in the right panels, where $\mu_B$ is the Bohr magneton.
    Here, $I_0 = N_0 e WD^2/L^3$, $W = 10L$, $L = 10\xi = \SI{400}{\nano\meter}$, $\delta/\Delta = 0.02$, $\Delta = \SI{1.5}{\milli\electronvolt}$ and $B_x = 0$.
  The blue line shows the analytical solution, as found by \cref{eq:currSol}, and the red line with dots shows the numerical solution found by solving the full Usadel equation.}%
  \label{fig:combined}
\end{figure}

Finally, \cref{fig:combined} shows the critical current for combinations of short and long ripples, both with and without a non-zero exchange field.
The height distribution is in this case given by
\begin{equation}
  \eta = \SI{1}{\nano\meter}\times\cos(2\pi\frac y W) + A\cos(n\pi\frac x L)\cos(n\pi\frac y L),
  \label{eq:combined}
\end{equation}
where $n = 4$ for ``hBN'' and ``Large'' and $n=10$ for ``SiO$_2$''.
The amplitude of the short ripples are $A = \SI{0.1}{\nano\meter}$, $A = \SI{0.5}{\nano\meter}$, $A = \SI{2}{\nano\meter}$ for ``hBN'', ``SiO$_2$''  and ``Large'', respectively.
The values for ``SiO$_2$'' and ``hBN'' are chosen such that the short ripple sizes correspond to the observed values for SiO$_2$ and hBN~\cite{xue2011}.
The values for ``Large'' are chosen such that amplitudes of the short ripples are twice as large as the long ripples.
In this case the orthogonal component of the magnetic field is much too high for the analytical solution to give accurate results.

Since $n$ is even, the analytical solution given by \cref{eq:currSol} is equal for the three cases.
The only difference is the magnitude of the additional magnetic flux density that comes from the short ripples.
As mentioned above, we should expect a faster decay for larger and faster ripples. This is indeed also what we observe from the numerical results in \cref{fig:combined}.
Interestingly, the location of the roots is not substantially altered by the short ripples. 
Even in the lowermost panels, where the short ripples are twice as large as the long ripples, the roots in the numerical solution occurs not far from the values of $B_y$ predicted by the analytical solution, even if the amplitude decays much faster.

Long ripples, as we have seen, can give rise to interference effects that produce oscillations and possibly roots in the critical current.
Short ripples, on the other hand, increase the magnetic flux density and can lead to a substantial magnetic depairing effect, which manifest as a rapidly decaying critical current.
Hence, the combined effect of short and long ripples can yield a rapidly decaying critical current with zeros, much like what one would expect from a ferromagnet undergoing a \mbox{0-$\pi$} transition.

From \cref{fig:combined} it can also be observed that the deviation between the analytical and numerical solutions is smaller when the exchange field is non-zero.
This is as expected, since neglecting the contribution from $B$ in \cref{eq:usadelWithMag} is a better approximation when the self-energy $\hat\Sigma$ is larger.
Note that in \cref{fig:combined}, the presence of the exchange field induces a \mbox{0-$\pi$} transition around $B_y = \SI{1}{\tesla}$, which manifests as a root in the critical current.

Since the ripples and the exchange field both give rise to oscillating and decaying critical currents, it is useful determine whether a \mbox{0-$\pi$} transition is expected to occur before or after a possible zero in the critical current coming from ripples.
The exact values of $B_y$ at which these events take place will in general depend on several parameters, but we can give some order of magnitude estimates based on \cref{eq:currSol} and numerical simulations.
The first \mbox{0-$\pi$} transition typically occurs around $h = 2\thouless$, but can occur at larger values if the inelastic scattering time, $1/\delta$, is small.
Inserting the definition of the Thouless energy, $\thouless$, and using that $h = \mu_B B_y$, this means that the Zeeman driven \mbox{0-$\pi$} transition occurs at
\begin{equation}
  B_y \approx \frac{2D}{\mu_B L^2}.
  \label{eq:magExc}
\end{equation}

\Cref{eq:magReq} gives a minimal value for $B_y$ at which a zero can be produced from the interference effect that is due to ripples.
In the numerical result presented here, we see that the zero occurs for a value of $B_y$ that is about three times larger. 
An order of magnitude estimate is that, for long ripples with peak-to-peak height of $\eta_0$, the first zero in the critical current can occur at around
\begin{equation}
  B_y \approx \frac{\Phi_0}{\eta_0 L}.
  \label{eq:magRip}
\end{equation}
Notice that \cref{eq:magExc,eq:magRip} scales differently with junction length $L$.
Therefore, it is more plausible that an observed zero in the critical current correspond to a \mbox{0-$\pi$} transition when the junction is long.
Alternatively, one could try to limit the presence of variations in the $y$-direction that are longer than $L$ by making $W \ll L$.

From \cref{fig:cos1_2,fig:cos1_5,fig:cos1_10} we also observe that the ripples can substantially alter the Fraunhofer lobe structure found when varying $B_\perp$, while \cref{fig:uniform} shows that the Fraunhofer pattern is unaltered when the effect of ripples is negligible.
Hence, investigating how $I_c$ depend on $B_\perp$ could also be useful when identifying \mbox{0-$\pi$} transitions.
As long as the junction width and diffusivity are approximately constant, a \mbox{0-$\pi$} transition will give rise to a vanishing critical current for all values of out-of-plane magnetic flux densities $B_\perp$.
If in addition the effect of ripples is small, one should expect that the critical current as a function of $B_\perp$ is a Fraunhofer pattern at any constant value of the in-plane magnetic field $B_\parallel$.
Accordingly, determining whether the minima in critical current as a function of $B_\perp$ and $B_\parallel$ are straight lines, as in \cref{fig:uniform}, or curved, as in \cref{fig:cos1_2,fig:cos1_5,fig:cos1_10}, can give clues as to whether ripples are important.
If ripples are important, \cref{eq:fourierVers} could give insight to their structure.

\section{Conclusion}%
\label{sec:discussion}
We have solved the Usadel equation analytically in the presence of an exchange field and an arbitrary magnetic field distribution, under the assumption of a weak proximity effect and a weak magnetic field.
The solution has been applied to SGS-junctions with the combined Zeeman effect and orbital effect coming from an in-plane magnetic field.
Deviations from the analytical solution at large magnetic fields have been studied numerically.
We find that the orbital effect that results from a curvature in the graphene can produce a critical current response that is similar to what one would get by increasing the exchange field.
Slow variations in the graphene height distributions give rise to interference effects that produce oscillations in the critical current, while rapid variations cause larger orbital depairing effects that lead to a faster critical current decay rate.

Since both the Zeeman splitting and orbital effects in rippled graphene can cause similar behaviour, extra care must be taken when identifying possible \mbox{0-$\pi$} transitions. 
The interference effect from ripples is reduced if the width of the junction is much smaller than the length.
In addition to reducing the relative effect of ripples compared to the Zeeman splitting, which is achieved by increasing the length of the junction and minimizing the height variations, it could also be useful to look at how the critical current varies with a perpendicular magnetic field.
The effect of ripples, if present, will then typically alter the Fraunhofer pattern observed at zero in-plane magnetic field.
Because slow height variations are difficult to detect using surface probe techniques, we suggest the use of parallel magnetic field as a means to probe the presence of such variations.
% Slow height variations are difficult to detect using surface probe techniques. We therefore suggest the use of parallel magnetic field as a means to probe the presence of such variations

% Slow height variations are difficult to detect using surface probe techniques. We therefore suggest the use of parallel magnetic field as a means to probe the presence of such variations

%-------------------------------------------------------------------------------%
%                                ACKNOWLEDGEMENTS                               %
%-------------------------------------------------------------------------------%
% Fakesection: Acknowledgements
\begin{acknowledgments}
This work was supported by the Research Council of Norway through grant 240806, and its Centres of Excellence funding scheme grant 262633 ``\emph{QuSpin}''. J. L. and M. A. also acknowledge funding from the NV-faculty at the Norwegian University of Science and Technology. 
H.S. is funded by a European Research Council Starting Grant (No. 637298, TUNNEL), and Israeli Science Foundation grant 861/19. T.D. and A.Z. are grateful to the Azrieli Foundation for Azrieli Fellowships.
\end{acknowledgments}

%-------------------------------------------------------------------------------%
%                                  TODOS                                        %
%-------------------------------------------------------------------------------%
% \newpage
% \listoftodos[Todos]

%-------------------------------------------------------------------------------%
%                                  BIBLIOGRAPHY                                 %
%-------------------------------------------------------------------------------%

% Fakesection: Bibliogaphy
\newpage
\bibliography{bibliography}

\end{document}